\documentstyle[12pt,aaspp4]{article}
\begin{document}

\title{High-Redshift Quasars Found in Sloan Digital
Sky Survey Commissioning Data II:
The Spring Equatorial Stripe$^1$}

\author{Xiaohui Fan\altaffilmark{\ref{Princeton}},
Michael A. Strauss\altaffilmark{\ref{Princeton}},
Donald P. Schneider\altaffilmark{\ref{PennState}},
James E. Gunn\altaffilmark{\ref{Princeton}},
Robert H. Lupton\altaffilmark{\ref{Princeton}},
Scott F. Anderson\altaffilmark{\ref{Washington}},
Wolfgang Voges\altaffilmark{\ref{MPI}},
Bruce Margon\altaffilmark{\ref{Washington}},
James Annis\altaffilmark{\ref{Fermilab}},
Neta A. Bahcall\altaffilmark{\ref{Princeton}},
J. Brinkmann\altaffilmark{\ref{APO}},
Robert J. Brunner\altaffilmark{\ref{JHU},\ref{Caltech}},
Michael A. Carr\altaffilmark{\ref{Princeton}},
Istv\'an Csabai\altaffilmark{\ref{JHU},\ref{Eotvos}},
Mamoru Doi\altaffilmark{\ref{UTokyo}},
Joshua A. Frieman\altaffilmark{\ref{Fermilab},\ref{Chicago}},
Masataka Fukugita\altaffilmark{\ref{CosmicRay},\ref{IAS}},
G. S. Hennessy\altaffilmark{\ref{USNO}},
Robert B. Hindsley\altaffilmark{\ref{USNO}},
\v{Z}eljko Ivezi\'{c}\altaffilmark{\ref{Princeton}},
G. R. Knapp\altaffilmark{\ref{Princeton}},
D. Q. Lamb\altaffilmark{\ref{Chicago}},
Timothy A. McKay\altaffilmark{\ref{Michigan}},
Jeffrey A. Munn\altaffilmark{\ref{Flagstaff}},
Heidi Jo Newberg\altaffilmark{\ref{Fermilab}},
A. George Pauls\altaffilmark{\ref{Princeton}},
Jeffrey R. Pier\altaffilmark{\ref{Flagstaff}},
Ron Rechenmacher\altaffilmark{\ref{Fermilab}},
Gordon T. Richards\altaffilmark{\ref{Chicago}},
Constance M. Rockosi\altaffilmark{\ref{Chicago}},
Chris Stoughton\altaffilmark{\ref{Fermilab}},
Alexander S. Szalay\altaffilmark{\ref{JHU}},
Aniruddha R. Thakar\altaffilmark{\ref{JHU}},
Douglas L. Tucker\altaffilmark{\ref{Fermilab}},
Patrick Waddell\altaffilmark{\ref{Washington}},
Donald G. York\altaffilmark{\ref{Chicago}}
}

\altaffiltext{1}{Based on observations obtained with the
Sloan Digital Sky Survey, and with the Apache Point Observatory
3.5-meter telescope, which is owned and operated by the Astrophysical
Research Consortium.}
\newcounter{address}
\setcounter{address}{2}
\altaffiltext{\theaddress}{Princeton University Observatory, Princeton, NJ 08544
\label{Princeton}}
\addtocounter{address}{1}
\altaffiltext{\theaddress}{Department of Astronomy and Astrophysics,
The Pennsylvania State University,
University Park, PA 16802
\label{PennState}}
\addtocounter{address}{1}
\altaffiltext{\theaddress}{University of Washington, Department of Astronomy,
Box 351580, Seattle, WA 98195
\label{Washington}}
\addtocounter{address}{1}
\altaffiltext{\theaddress}{Max-Planck-Institut f\"{u}r extraterrestrische Physik, Postfach 1603, 85750 Garching, Germany 
\label{MPI}}
\addtocounter{address}{1}
\altaffiltext{\theaddress}{Fermi National Accelerator Laboratory, P.O. Box 500,
Batavia, IL 60510
\label{Fermilab}}
\addtocounter{address}{1}
\altaffiltext{\theaddress}{Apache Point Observatory, P.O. Box 59,
Sunspot, NM 88349-0059
\label{APO}}
\addtocounter{address}{1}
\altaffiltext{\theaddress}{
Department of Physics and Astronomy, The Johns Hopkins University,
   3701 San Martin Drive, Baltimore, MD 21218, USA
\label{JHU}}
\addtocounter{address}{1}
\altaffiltext{\theaddress}{
Department of Astronomy, California Institute of Technology,
Pasadena, CA 91125
\label{Caltech}
}
\addtocounter{address}{1}
\altaffiltext{\theaddress}{Department of Physics of Complex Systems,
E\"otv\"os University,
   P\'azm\'any P\'eter s\'et\'any 1/A, Budapest, H-1117, Hungary
\label{Eotvos}
}
\addtocounter{address}{1}
\altaffiltext{\theaddress}{Department of Astronomy and Research Center for the Early
Universe, School of Science, University of Tokyo, Hongo, Bunkyo,
Tokyo, 113-0033 Japan
\label{UTokyo}}
\addtocounter{address}{1}
\altaffiltext{\theaddress}{University of Chicago, Astronomy \& Astrophysics
Center, 5640 S. Ellis Ave., Chicago, IL 60637
\label{Chicago}}
\addtocounter{address}{1}
\altaffiltext{\theaddress}{Institute for Cosmic Ray Research, University of
Tokyo, Midori, Tanashi, Tokyo 188-8502, Japan
\label{CosmicRay}}
\addtocounter{address}{1}
\altaffiltext{\theaddress}{Institute for Advanced Study, Olden Lane,
Princeton, NJ 08540
\label{IAS}}
\addtocounter{address}{1}
\altaffiltext{\theaddress}{U.S. Naval Observatory,
3450 Massachusetts Ave., NW,
Washington, DC  20392-5420
\label{USNO}}
\addtocounter{address}{1}
\altaffiltext{\theaddress}{University of Michigan, Department of Physics,
        500 East University, Ann Arbor, MI 48109
\label{Michigan}}
\addtocounter{address}{1}
\altaffiltext{\theaddress}{U.S. Naval Observatory, Flagstaff Station,
P.O. Box 1149,
Flagstaff, AZ  86002-1149
\label{Flagstaff}}

\begin{abstract}
This is the second paper in a series aimed at
finding high-redshift quasars from five-color ($u'g'r'i'z'$)
imaging data taken along the Celestial Equator by 
the Sloan Digital Sky Survey (SDSS)
during its commissioning phase. 
In this paper, we present 22 high-redshift quasars ($z>3.6$)
discovered from  $\sim 250$ deg$^2$ of data in the spring
Equatorial Stripe, plus photometry for two previously known
high-redshift quasars in the same region of sky.  
Our success rate of identifying high-redshift quasars is 68\%. 
Five of the newly discovered quasars have redshifts higher than 4.6
($z=4.62$, 4.69, 4.70, 4.92 and 5.03).
All the quasars have $i^* < 20.2$ with absolute magnitude
$\rm -28.8 < M_{B} < -26.1 $ ($h=0.5$, $q_{0}=0.5$).
Several of the quasars show unusual emission and absorption features in
their spectra, including an object at $z=4.62$ without detectable emission
lines, and a Broad Absorption Line (BAL)  quasar at $z=4.92$.

\end{abstract}

\section{Introduction}
This paper is the second in a series presenting high-redshift
quasars selected from the commissioning data of the 
Sloan Digital Sky Survey (SDSS\footnote{\texttt{http://www.astro.princeton.edu/PBOOK/welcome.htm}}, \cite{York99}).
In Paper I (\cite{paper1}), we presented the discovery of 15 quasars
at $z>3.6$ selected from two SDSS photometric runs covering
$\approx 140$ deg$^2$ in the Southern Galactic cap
along the Celestial Equator observed in Fall  1998.  
In this paper, we describe observations
of quasar candidates selected in a similar manner from 250 deg$^2$ of
SDSS imaging data in the Northern Galactic cap, again along the
Celestial Equator, observed in Spring  1999.  
The scientific objectives, photometric data reduction,
selection criteria and spectroscopic observation procedures are
described in Paper I, and will be outlined only briefly here.

We have not yet observed all the quasar candidates spectroscopically, so
the objects described in these two papers do not form a complete
sample.  We will present the complete sample of
high-redshift quasars found in the Equatorial stripe,
and derive the quasar luminosity function at high redshift, in 
subsequent papers. 

We describe the photometric observations and selection of quasar
candidates briefly in \S 2. The spectra of 22 new high-redshift
quasars are presented in \S 3. 

\section{Photometric Observations and Quasar Selection}

The SDSS telescope (\cite{Siegmund99}\footnote{see also {\tt
http://www.astro.princeton.edu/PBOOK/telescop/telescop.htm}}), 
imaging camera (\cite{Gunnetal}), 
and photometric data reduction (\cite{Lupton99b}\footnote{see also 
{\tt http://www.astro.princeton.edu/PBOOK/datasys/datasys.htm}}) 
are described in Paper I.  Briefly, the telescope, located
at Apache Point Observatory in southeastern New Mexico, has a 2.5m
primary mirror and a wide, essentially distortion-free field.  The
imaging camera contains thirty $2048 \times 2048$ photometric CCDs, which
simultaneously observe 6 parallel $13'$ wide swaths, or {\em scanlines}
of the sky, in 5 broad filters ($u'$, $g'$, $r'$, $i'$, and $z'$) covering
the entire optical band from the atmospheric cutoff in the blue to
the silicon sensitivity cutoff in the red (\cite{F96}). 
The photometric data are taken in time-delay and integrate (TDI, or
``drift-scan'') mode at sidereal rate; 
a given point on the sky passes through each of the five filters in
succession.
The total integration time per filter is 54.1 seconds.
The data were calibrated photometrically by observing secondary standards
in the survey area using a (now decommissioned) 60cm telescope at
Apache Point Observatory and the US Naval Observatory's 1m telescope.
The photometric calibration used in this paper
is only accurate to 5--10\%, due to systematics in the shape of the
point spread function across individual CCDs, and the fact that the
primary standard star network had not yet been finalized at the time of 
these observations.  
This situation will be improved to the survey requirement of 2\%
in future papers in this series. 
Thus as in Paper I, we will denote the preliminary SDSS magnitudes
presented here as $u^*$, $g^*$, $r^*$, $i^*$ and $z^*$, rather than
the notation $u'$, $g'$, $r'$, $i'$ and $z'$ that will be used for the
final SDSS photometry.

In this paper, we select quasar candidates from four SDSS imaging runs
in the Northern Galactic Cap. 
The data were acquired with the telescope parked at the Celestial
Equator. 
Details of the photometric runs are
summarized in Table 1. 
Two interleaved SDSS scans, or the Northern and Southern {\em strips}, 
form a filled {\em stripe} 2.5 degrees wide in declination, centered on the
Celestial Equator.
Run 77 and 745 cover the Northern strip of the Equatorial Stripe,
while Runs 85 and 752 cover the Southern strip. 
Runs 77 and 745, and runs 85 and 752 have considerable overlap, 
but candidates were selected only on the catalogs based on individual runs.
The total stripe is roughly 7 hours long and covers a total area of about 250
deg$^2$ at Galactic latitude in the range $25^\circ < b < 63^\circ$. 
All four nights were photometric, with seeing conditions varying from $1.3''$ to
worse than $2''$. 

The data are processed by a series of automated pipelines to carry out
astrometric and photometric measurements (c.f. Paper I, and references
therein).
The final object catalog includes roughly 5 million objects in total. 
The limiting magnitudes are similar to those of Paper I, roughly
22.3, 22.6, 22.7, 22.4 and 20.5 in $u^*$, $g^*$, $r^*$, $i^*$ and $z^*$,
respectively.  
Figure 1 presents the color-color diagrams from Run 752 
for all stellar sources at $i^* < 20.2$. 
The inner parts of the diagrams are shown as contours, linearly spaced
in density of stars per unit area in color-color space.  
As in Paper I, a source is plotted only if it is detected in all three
of the relevant bands at more than 5$\sigma$.  In addition, objects that are flagged as
saturated, lying on the bleed trail of a saturated star, overlapping the
edge of the image boundary, or showing other indications of possible problems
in the photometric measurement, are rejected.  
The median tracks of quasar colors as a function of redshift, as well as the
locations of low-redshift ($z<2.5$) quasars, hot white dwarfs and A stars, all
from the simulation of \cite{Fan99}, are also plotted in Figure 1.

High-redshift quasar candidates were selected using color cuts similar
to those presented in Paper I. 
Because of the uncertainties in the photometric zeropoints,
we found that the stellar locus shifted by of order 0.05 mag in the
color-color diagrams between the Fall (Paper I) and Spring observations.
We adjust the color cuts presented in Paper I according to these shifts.
Final color cuts of the complete sample will be presented 
in a later paper with the final photometric calibrations.
The color selection criteria used in this paper are as follows:

1. $gri$ candidates, selected principally from the $g^*-r^*, r^*-i^*$ diagram:

\begin{equation}
 \begin{array}{l}
        (a)\ i^* < 20 \\
        (b)\ u^* - g^* > 2.0 \mbox{ or } u^* > 22.3 \\
        (c)\ g^* - r^* > 1.0 \\
        (d)\ r^* - i^* < 0.08 + 0.42 (g^* - r^* -0.96) \mbox{ or } g^*- r^* >2.26 \\
        (e)\ i^* - z^* < 0.25
        \end{array}
\end{equation}

2. $riz$ candidates, selected principally from the $r^*-i^*, i^*-z^*$ diagram:

\begin{equation}
\begin{array}{l}
        (a)\ i^* < 20.2  \\
        (b)\ u^* > 22.3 \\
        (c)\ g^* > 22.6 \\
        (d)\ r^* - i^* > 0.8 \\
        (e)\ i^* - z^* < 0.47 (r^* - i^* - 0.68)
        \end{array}
\end{equation}

The intersections of those color cuts with the $g^* - r^*, r^*-i^*$ and
$r^*-i^*,i^*-z^*$ diagrams are illustrated in Figure 1.  A total of
80  $gri$ and $riz$ candidates that have colors consistent with quasars at
$z>3.6$  and $i^* < 20.0$ were selected from the catalog.  Several other
$riz$ candidates ($z>4.6$) at $i^* < 20.2$  were
also selected and observed.  Two of the candidates, SDSSp
J105320.43--001649.3 and SDSSp J111246.30+004957.5 (see Table 2; for
naming convention, see \S 3), are the previously known quasars
BRI1050--0000 ($z=4.29$, \cite{APM}) and BRI1110+0106 ($z=3.92$,
\cite{Smith94}).  Those two objects are the only quasars with $z>3.6$
in the area covered in the NED database\footnote{The NASA/IPAC
Extragalactic Database (NED) is operated by the Jet Propulsion
Laboratory, California Institute of Technology, under contract with the
National Aeronautics and Space Administration.}.

\section{Spectroscopic Observations}

Spectra of 32 high-redshift quasar candidates from the Equatorial
Stripe were obtained with the ARC 3.5m telescope of the Apache Point 
Observatory, using the Double Imaging Spectrograph (DIS), during a
number of nights from March to May 1999. 
Exposure times of these candidates range from
600 seconds for the brightest ($i^* \sim 17$) candidate to 5400 seconds for 
the faintest candidates. The instrument and spectral data reduction procedures
were described in Paper I.
The final spectra extend from 4000 \AA\ to 10000 \AA, 
with a spectral resolution of 12 \AA\ in the blue and 14 \AA\ in the
red.  They have been flux calibrated and corrected for telluric
absorption with observations of F subdwarfs (\cite{OG83}, \cite{Oke90}). 
Twenty-one of the candidates are identified to be high-redshift quasars at $z>3.6$.
In particular, five of the candidates are quasars at $z>4.6$, with redshifts
of 4.62, 4.69, 4.70, 4.92 and 5.03, respectively.
Two of these  objects have very unusual spectra.\\
SDSSp J153259.96--003944.1 is identified as a quasar without detectable emission
lines at $z=4.62$. Its optical, radio, and polarization properties are reported
in a separate paper (\cite{bllac}).
SDSSp J160501.21--011220.0 is a Broad Absorption Line (BAL) quasar at
a redshift of 4.92 with two BAL systems, and is described further
below. 
One additional object, SDSSp J130348.94+002010.4, was also observed.
It has redder $g^*-r^*$ and $r^*-i^*$ colors than required by our color   
selection criteria (eq. 1), but is identified as a BAL quasar at $z=3.64$.
We present its spectrum below but will not include it in the future
statistical analyses.

Table 2 gives the position and SDSS photometry, 
the redshift of each confirmed SDSS
quasar and the photometric run from which it was selected. 
For the objects in the overlap region between runs, 
only the results from the runs from which they
are selected are listed. 
None of these quasars show magnitude differences of more than 0.2 mag
in the high signal-to-noise passbands between runs.
We also include the SDSS measurements of the two previously known $z>3.6$
quasars in Table 2. 
The naming convention for the SDSS sources is SDSSp J HHMMSS.SS$\pm$DDMMSS.S,
where ``p'' stands for the preliminary SDSS astrometry, and the positions
are expressed in J2000.0 coordinates.
The preliminary SDSS astrometry is accurate to better than $0.2''$
in each axis. The photometry is expressed in asinh magnitudes 
(\cite{Luptitude}, see also Paper I); this magnitude system approaches normal
logarithmic magnitudes at high signal-to-noise ratio, but becomes a
linear flux scale for low signal-to-noise ratio, even for slightly
negative fluxes. The photometric errors given are 
statistical in nature and do not include the 
systematic errors due to PSF variation across the field and uncertainties in the photometric zeropoint.
Positions of the 24 confirmed SDSS quasars on the color-color diagrams are
plotted on Figure 1. 
Finding charts of all objects in Table 2 are given in Figure 2. 
They  are $200'' \times 200''$ $i'$ band SDSS images with 
an effective exposure time of 54.1 seconds.

We matched the positions of quasars in Table 2 against radio surveys.
Twenty  of them are in the region covered by the FIRST survey
(\cite{FIRST}). Three of them have FIRST counterparts at 20 cm at the 1 mJy level
(with positional matches better than $1''$).
Two new SDSS quasars, SDSSp J123503.04--000331.8 ($z=4.69$) and
SDSSp J141205.78--010152.6 ($z=3.73$), correspond to
FIRST sources of 18.4 and 4.3 mJy at 20 cm, respectively.
One of the previously known quasars, SDSSp J105320.43--001649.3 (BRI1050-0000, $z=4.
29$),
is also a FIRST source of 13.8 mJy.
The three quasars with RA $> 16^h$ are not covered by the FIRST survey.
We matched them against the NVSS survey (\cite{NVSS});
none of them has an NVSS counterpart at 20cm at the 2.5 mJy level.

We have similarly cross-correlated the list against the ROSAT full-sky
pixel images (\cite{ROSAT}); none of these objects were detected, implying a typical
3-$\sigma$ upper limit of $\rm 3 \times 10^{-13} erg\ cm^{-2} s^{-1}$ in the
0.1 -- 2.4 keV band.
This result is not unexpected; only a few 
$z>4$ quasars have an observed X-ray flux above this value
(e.g. \cite{Fabian97}, \cite{Moran97}), and the typical
X-ray flux from known optically selected $z>4$ quasars is a factor of
four or more lower than our limit (\cite{Brandt99}).
We expect positive ROSAT X-ray detections of a substantial
fraction of the somewhat brighter, lower-redshift quasars in the SDSS.

Among the 34 observed candidates which satisfy Equations 1 and 2 (including the two previously known
quasars), 23 are confirmed as quasars at $z>3.6$, a success rate of
68\%, similar to the success rate we reported in Paper I.  Ten of the eleven 
non-quasars are faint late type stars, which are typical
contaminants of high-redshift quasar searches.  
One of the candidates,
however, has spectral features that we have not yet been able to
identify; we will present its spectrum and discuss possible
explanations in a separate paper.

Figure 3 presents our spectra of the 22 new SDSS quasars. 
In Figure 3, we place the spectra on an absolute flux scale
(to compensate for the uncertainties due to non-photometric conditions
and variable seeing during the night) by forcing the synthetic
$i^*$ magnitudes from the spectra to be the same as
the SDSS photometric measurements.
The synthetic and photometric measurements agree within $\sim 0.1$ mag
for the objects observed in photometric nights.
This scatter is due to both the uncertainties in the SDSS
photometric zeropoints (5 -- 10\%, see \S 2) 
and the spectroscopic observations.
Therefore, the absolute flux scale in Figure 3 is only accurate to
$\sim 0.1$ mag.
 
The emission line properties of the quasars are listed in Table 3.
Central wavelengths and rest frame equivalent widths of five major 
emission lines are measured following the procedures of Paper I.
Table 4 gives the continuum properties of the quasars.  As in Paper I,
redshifts are determined from all emission lines redward of
Ly$\alpha$; Ly$\alpha$ itself is not used, due to absorption from the
Ly$\alpha$ forest on its blue side.  
The AB magnitude of the quasar continuum at 1450 \AA\ (rest frame),
$AB_{1450}$, is determined by the average flux in the continuum
window from 1425 \AA\ to 1475 \AA\ (rest frame) 
from the spectra in Figure 3.
The absolute magnitude $M_{B}$ is determined by assuming
a cosmology of $q_{0}=0.5$, $h=0.5$, and a power law index of --0.5
following Schneider, Schmidt \& Gunn (1991).
The $\rm AB_{1450}$ and $M_{B}$ magnitudes are corrected for 
Galactic extinction using the reddening map of \cite{Schlegel98};
these values are accurate to $\sim 0.1$ mag, due to the uncertainties
in the absolute flux scale (see above).
The emission and absorption properties of most of these quasars are 
very similar to those of
other quasars at similar redshift (c.f. \cite{SSG89}, 1991, 1997,
\cite{Warren91},
\cite{Kennefick95}, \cite{APM}, Paper I), with some interesting
exceptions, which we now discuss. 

\subsection{Notes on Individual Objects}

\noindent {\bf SDSSp J112253.51+005329.8} ($z=4.57$).
A number of absorption systems are present in the spectrum.
In particular, the peak of the Ly$\alpha$ emission line is 
self-absorbed by several lines. 

\noindent {\bf SDSSp J120441.73--002149.6} ($z=5.03$). 
This is the second quasar found at $z \gtrsim 5$ in our survey. 
As was the case for SDSSp J033829.31--002156.3 ($z=5.00$, Paper I), 
the determination of an accurate
redshift is not straightforward.
A straight Gaussian fit to the Ly$\alpha$ emission line yields
a redshift of 5.14. 
The Si$\,$IV and C$\,$IV emissions are affected by atmospheric absorption;
they give a consistent redshift of about 5.03. 
We therefore adopt a redshift of 5.03 $\pm$ 0.05 for this quasar.
For these two  $z \gtrsim 5$ quasars, the differences between the redshifts
of the Ly$\alpha$ lines and that of the Si$\,$IV and C$\,$IV lines are about 0.1,
much larger than other quasars at $z>4$ (\cite{SSG91a}). 
It is not clear whether the Lyman $\alpha$
absorption at the blue wing of the Lyman $\alpha$ emission line can
fully account for this large redshift discrepancy.
The Ly$\alpha$ line profile is also more symmetric than that of
other high-redshift quasars.
This object is relatively bright  at $i^*=19.3$, suitable for high 
signal-to-noise ratio studies of its emission line profiles. 

\noindent {\bf SDSSp J130348.94+002010.4} ($z=3.64$). This BAL quasar shows absorption 
troughs shortward of $z\approx3.64$ (as measured by the emission peaks) in 
virtually all sampled strong emission lines, extending blueward by up to 
$\sim 13,000$ km~s$^{-1}$ (e.g., for C\,IV). Although this object is a marked 
outlier in SDSS colors, it is redder in $g^*-r^*$ and $i^{*}-z^{*}$ (perhaps 
because of the BALs) than the great majority of other known high-redshift 
quasars, and does not satisfy our formal selection 
criteria (eq. 1). 

\noindent {\bf SDSSp J141205.78--010152.6} ($z=3.73$).
This object possesses an interesting absorption system at $z=3.62$ ($v = - 7000$ km s$^{-1}$)
with absorption lines of C$\,$IV (7158\AA), N$\,$V (5729\AA), Ly$\alpha$
(5619\AA) and Ly$\beta$+O$\,$VI (4763\AA).
The C$\,$IV absorption has a FWHM of $\rm \sim 1300\ km\ s^{-1}$.
This system is probably also responsible for the peculiar shape of
the Lyman Limit System.
High resolution observations are needed to determine
the nature of this absorption system,
especially whether it is  a so-called ``mini-BAL'' system
(defined as  the velocity span of the absorption profile being
narrower than 2000 $\rm km\ s^{-1}$; \cite{Barlow97}, \cite{Churchill99}).

\noindent {\bf SDSSp J141315.36+000032.1} ($z=4.08$).
This object was observed under poor weather conditions, and the
spectrum has a very low signal-to-noise ratio.
However, the Ly$\alpha$ and C$\,$IV emission lines are clearly detected,
and give consistent redshifts.

\noindent {\bf SDSSp J141332.35--004909.7} ($z=4.14$).
This is another mini-BAL quasar candidate.
The C$\,$IV trough has a FWHM of $\rm \sim 1400\ km\ s^{-1}$.
An accurate center and equivalent width of the Ly$\alpha$ line
cannot be measured due to the presence of the BAL trough of N$\,$V,
which appears blueward of the Ly$\alpha$ emission.


\noindent {\bf SDSSp J151618.44--000544.3} ($z=3.70$).
The spectrum of this object shows a strong damped Ly$\alpha$ system candidate at $z=3.03$ with
rest-frame equivalent width of $\sim 38$ \AA. 
In comparison, the strongest known damped Ly$\alpha$ system has
a rest-frame  equivalent width of 41 \AA\ 
(\cite{Wolfe86}).
The spectrum at $\lambda > 7500$\AA\ is affected by a CCD defect and
is not plotted.

\noindent {\bf SDSSp J153259.96--003944.1} ($z=4.62$).
This is a unique object. 
Its spectrum features two breaks at 6800 \AA\ and
5100 \AA, respectively. 
We interpret the two breaks as the onset of the Lyman $\alpha$ forest and 
a Lyman Limit System respectively, giving a consistent redshift of $z=4.62$.
The redshifts of onset of the Lyman $\alpha$ forest and Lyman Limit System are typically
close to the emission line redshift for high-redshift quasars
(e.g., \cite{SSG91b}, \cite{APM});
we therefore adopt a redshift of 4.62 $\pm$ 0.04 for this object.
However, this object has no detected emission lines. 
We discuss high signal-to-noise ratio Keck spectroscopy, VLA radio observations and
optical broad-band polarimetry of this object in a separate paper (\cite{bllac}).

\noindent {\bf SDSSp J160501.21--011220.0} ($z=4.92$).
This object is the highest redshift BAL quasar yet discovered.
It has two BAL systems in its spectrum, one at $z=4.86$,
the other at $z=4.69$.
The absorption due to the BAL systems is detected 
for the Ly$\alpha$, N$\,$V, Si$\,$IV and C$\,$IV lines.
Because of the BALs, an emission line redshift cannot be measured
accurately.
We adopt a redshift of $4.92\pm 0.05$, as measured from the
peaks of the N$\,$V, O$\,$I and Si$\,$IV lines. 
In Figure 4, we show the absorption systems for each line,
aligned in the rest-frame velocity of the quasar.
The two BALs are at relative velocities of 3000 km $\rm s^{-1}$ and
11,700 km $\rm s^{-1}$, respectively.
This is an ideal object for further spectroscopic observations
to study the BAL phenomenon at very high redshift.	
The presence of the BALs affects the broad-band colors, 
resulting in $r^*-i^* \sim 3$, compared to $r^*-i^* < 2$
for other $z \sim 5$ quasars (Figure 1c).

\noindent {\bf SDSSp J162116.91--004251.1} ($z=3.70$).
This is a very bright high-redshift quasar ($i^*=17.23$, $\rm
M_{B}=-28.81$), the most luminous we have yet found.  The
signal-to-noise ratio of this spectrum is high, allowing
identification of several absorption lines.
It is particularly suitable for high-resolution studies 
of its absorption systems. 

\noindent {\bf SDSSp J165527.61--000619.2} ($z=3.99$).
This object has a very strong Ly$\alpha$ emission line; the
rest-frame equivalent width is 172 \AA, which is more than twice the
value of most quasars in this redshift range. 
The presence of this strong emission line affects the broad-band colors.
It has $g^*-r^* \sim 3$, compared to $g^*-r^* \sim 1.6$ for ordinary $z\sim 4$
quasars (Figure 1b).

\bigskip

The Sloan Digital Sky Survey (SDSS) is a joint project of the
University of Chicago, Fermilab, the Institute for Advanced Study, the
Japan Participation Group, The Johns Hopkins University, the
Max-Planck-Institute for Astronomy, Princeton University, the United
States Naval Observatory, and the University of Washington.  Apache
Point Observatory, site of the SDSS, is operated by the Astrophysical
Research Consortium.  Funding for the project has been provided by the
Alfred P. Sloan Foundation, the SDSS member institutions, the National
Aeronautics and Space Administration, the
National Science Foundation, the U.S. Department of Energy, and the
Ministry of Education of Japan.  
The SDSS Web site is {\tt http://www.sdss.org/}.
XF and MAS acknowledge additional
support from Research Corporation, NSF grants AST96-16901, the
Princeton University Research Board, and an Advisory Council
Scholarship. DPS acknowledges the support of NSF grant AST95-09919 and
AST99-00703.
We thank Niel Brandt for useful discussions.
We thank Karen Gloria and Russet McMillan for their usual expert
help at the 3.5m telescope.

\newpage

\begin{deluxetable}{ccccc}
\tablenum{1}
\tablecolumns{5}
\tablewidth{0pc}
\tablecaption{Summary of SDSS Photometric Runs}
\tablehead
{
Run & Date & Strip & RA Range & seeing
}
\startdata
77  &  June 27 1998  & North & $14^h10^m - 17^h$ & $1.6''- 1.9''$ \\
85  &  June 28 1998  & South & $14^h20^m - 17^h$ & $1.3''- 2.0''$ \\
745 &  Mar  20 1999  & North & $10^h40^m - 16^h40^m$ & $1.3'' - 1.6''$\\
752 &  Mar  21 1999  & South & $9^h40^m - 16^h40^m$ & $1.3'' - 2.5''$ \\
\enddata
\end{deluxetable}

\begin{scriptsize}
\begin{deluxetable}{lccccccc}
\tablenum{2}
\tablecolumns{8}
\tablewidth{0pc}
\tablecaption{Positions and Photometry of SDSS High-redshift Quasars}
\tablehead
{
SDSS name & redshift & $u^*$ & $g^*$ & $r^*$ & $i^*$ & $z^*$ & run
}
\startdata
SDSSp J$105320.43-001649.3^a$ & 4.29 $\pm$ 0.01& 23.81 $\pm$ 0.43 & 21.75 $\pm$ 0.07 & 19.36 $\pm$ 0.02
& 19.32 $\pm$ 0.02 & 19.33 $\pm$ 0.07  & 752\\
SDSSp J$111246.30+004957.5^b$ & 3.92 $\pm$ 0.01 & 24.19 $\pm$ 0.36 & 20.09 $\pm$ 0.02 & 18.82 $\pm$ 0.01 & 18.69 $\pm$ 0.01 & 18.64 $\pm$ 0.04  & 752\\
SDSSp J$111401.48-005321.1$ & 4.58 $\pm$ 0.01 & 23.42 $\pm$ 0.23 & 22.94 $\pm$ 0.11 & 20.64 $\pm$ 0.03 & 19.60 $\pm$ 0.02 & 19.56 $\pm$ 0.06  & 745 \\
SDSSp J$112253.51+005329.8$ & 4.57 $\pm$ 0.02& 23.45 $\pm$ 0.46 & 22.78 $\pm$ 0.22 & 20.21 $\pm$ 0.03 & 19.11 $\pm$ 0.02 & 19.11 $\pm$ 0.07  & 752 \\
SDSSp J$120441.73-002149.6$ & 5.03 $\pm$ 0.05 & 22.29 $\pm$ 0.21 & 24.92 $\pm$ 0.01 & 20.82 $\pm$ 0.06 & 19.31 $\pm$ 0.02 & 19.12 $\pm$ 0.08 & 752\\ \\

SDSSp J$122600.68+005923.6$ & 4.25 $\pm$ 0.02 & 23.00 $\pm$ 0.32 & 21.26 $\pm$ 0.05 & 19.01 $\pm$ 0.01 & 18.91 $\pm$ 0.02 & 18.86 $\pm$ 0.05 & 752\\
SDSSp J$123503.04-000331.8$ & 4.69 $\pm$ 0.01 & 23.58 $\pm$ 0.23 & 23.85 $\pm$ 0.28 & 21.49 $\pm$ 0.06 & 20.11 $\pm$ 0.03 & 20.05 $\pm$ 0.09 & 745 \\
SDSSp J$130348.94+002010.5^c$ & 3.64 $\pm$ 0.03 & 23.22 $\pm$ 0.25 & 20.94 $\pm$ 0.03 & 18.93 $\pm$ 0.01 & 18.71 $\pm$ 0.01 & 18.25 $\pm$ 0.03 & 745 \\
SDSSp J$131052.52-005533.4$ & 4.14 $\pm$ 0.01 & 23.28 $\pm$ 0.49 & 20.85 $\pm$ 0.02 & 18.85 $\pm$ 0.01 & 18.82 $\pm$ 0.01 & 18.88 $\pm$ 0.03 & 745 \\
SDSSp J$132110.82+003821.7$ & 4.70 $\pm$ 0.01 & 23.67 $\pm$ 0.24 & 23.49 $\pm$ 0.19 & 21.40 $\pm$ 0.06 & 20.01 $\pm$ 0.03 & 20.17 $\pm$ 0.09  & 745 \\ \\

SDSSp J$141205.78-010152.6$ & 3.73 $\pm$ 0.02 & 24.09 $\pm$ 0.37 & 20.58 $\pm$ 0.02 & 19.39 $\pm$ 0.02 & 19.22 $\pm$ 0.02 & 19.02 $\pm$ 0.06 & 77\\
SDSSp J$141315.36+000032.1$ & 4.08 $\pm$ 0.02 & 23.61 $\pm$ 0.34 & 21.46 $\pm$ 0.05 & 19.74 $\pm$ 0.02 & 19.72 $\pm$ 0.03 & 19.77 $\pm$ 0.10 & 752 \\
SDSSp J$141332.35-004909.7$ & 4.14 $\pm$ 0.01 & 24.11 $\pm$ 1.62$^d$ & 21.11 $\pm$ 0.04 & 19.59 $\pm$ 0.02 & 19.29 $\pm$ 0.02 & 19.17 $\pm$ 0.06 & 752 \\
SDSSp J$142329.98+004138.4$ & 3.76 $\pm$ 0.01 & 23.54 $\pm$ 0.33 & 20.77 $\pm$ 0.03 & 19.55 $\pm$ 0.02 & 19.52$\pm$  0.03 & 19.56 $\pm$ 0.08 & 77 \\
SDSSp J$142647.82+002740.4$ & 3.69 $\pm$ 0.01 & 22.45 $\pm$ 0.13 & 20.55 $\pm$ 0.02 & 19.40 $\pm$ 0.01 & 19.34 $\pm$ 0.02 & 19.28 $\pm$ 0.05 & 85 \\ \\

SDSSp J$144428.67-012344.1$ & 4.16 $\pm$ 0.01 & 23.37 $\pm$ 0.35 & 21.24 $\pm$ 0.04 & 19.52 $\pm$ 0.02 & 19.47 $\pm$ 0.02 & 19.37 $\pm$ 0.07 & 85 \\
SDSSp J$144758.46-005055.4$ & 3.80 $\pm$ 0.01 & 23.50 $\pm$ 0.26 & 20.98 $\pm$ 0.03 & 19.58 $\pm$ 0.02 & 19.37 $\pm$ 0.02 & 19.14 $\pm$ 0.07 & 77\\
SDSSp J$151618.44-000544.3$ & 3.70 $\pm$ 0.01 & 24.10 $\pm$ 0.31 & 21.85 $\pm$ 0.08 & 20.18 $\pm$ 0.02 & 19.99 $\pm$ 0.03 & 19.90 $\pm$ 0.12 &  752\\
SDSSp J$152740.52-010602.6$ & 4.41 $\pm$ 0.02 & 23.98 $\pm$ 0.14 & 22.78 $\pm$ 0.13 & 20.46 $\pm$ 0.03 & 19.92 $\pm$ 0.03 & 19.67 $\pm$ 0.08 & 752\\ 
SDSSp J$153259.96-003944.1$ & 4.62 $\pm$ 0.04 & 23.74 $\pm$ 0.36 & 23.78 $\pm$ 0.30 & 21.15 $\pm$ 0.05 & 19.75 $\pm$ 0.03 & 19.55 $\pm$ 0.07 & 77 \\ \\

SDSSp J$160501.21-011220.0$ & 4.92 $\pm$ 0.05 & 24.08 $\pm$ 0.36 & 25.04 $\pm$ 0.49 & 22.50 $\pm$ 0.17 & 19.78 $\pm$ 0.03 & 19.92 $\pm$ 0.10 & 85 \\
SDSSp J$161926.87-011825.2$ & 3.84 $\pm$ 0.01 & 22.77 $\pm$ 0.24 & 21.80 $\pm$ 0.07 & 20.17 $\pm$ 0.03 & 19.93 $\pm$ 0.03 & 20.02 $\pm$ 0.13 & 85 \\
SDSSp J$162116.91-004251.1$ & 3.70 $\pm$ 0.01 & 22.09 $\pm$ 0.11 & 18.62 $\pm$ 1.09$^d$ & 17.28 $\pm$ 0.00 & 17.23 $\pm$ 0.00 & 17.26 $\pm$ 0.01 & 752 \\
SDSSp J$165527.61-000619.2$ & 3.99 $\pm$ 0.01 & 24.57 $\pm$ 0.21 & 22.92$\pm$  0.17 & 20.05 $\pm$ 0.02 & 20.16 $\pm$ 0.04 & 20.17 $\pm$ 0.15  & 77 \\
\enddata
\tablenotetext{}{Positions are in J2000.0 coordinates; 
asinh magnitudes (Lupton, Gunn \& Szalay 1999) are
quoted; errors are statistical only.  For reference, zero flux
corresponds to asinh magnitudes of 23.40, 24.22, 23.98, 23.51, and
21.83 in $u^*, g^*, r^*, i^*$, and $z^*$, respectively.}
\tablenotetext{a}{This is the previously known quasar BRI1050--0000 (Storrie-Lombardi
{\em et al.} 1996).}
\tablenotetext{b}{This is the previously known quasar BRI1110+0106 (Smith, Thompson \& Djorgovski 1994).} 
\tablenotetext{c}{This object does not satisfy the color selection criteria (eq. 1).}
\tablenotetext{d}{The large magnitude error indicates that the
magnitude is not to be trusted.}
\end{deluxetable}
\end{scriptsize}

\begin{scriptsize}
\begin{deluxetable}{crrrrr}
\tablenum{3}
\tablecolumns{6}
\tablewidth{0pc}
\tablecaption{Emission Line Properties of SDSS High-redshift Quasars}
\tablehead
{
quasar           &      O$\,$VI  & Ly$\alpha$ & O$\,$I+Si$\,$II & Si$\,$IV+O$\,$IV] & C$\,$IV   \\
                 & 1034   &  1216+1240 & 1306 & 1402  & 1549
}
\startdata
SDSSp J111401.48-005321.1 &               & 6841 $\pm$  4 
&               & 7822 $\pm$ 16 & 8642 $\pm$ 18 \\ 
&                 &  47.6 $\pm$  3.8 &                 
&  14.3 $\pm$  2.1 &  28.4 $\pm$  3.6 \\ 
SDSSp J112253.51+005329.8 &               & 6797 $\pm$  2 
& 7290 $\pm$  4 & 7804 $\pm$  6 & 8607 $\pm$  3 \\ 
&                 &  96.0 $\pm$  1.4 &   3.4 $\pm$  0.4 
 &  29.1 $\pm$  2.3 &  42.7 $\pm$  1.3 \\ 
SDSSp J120441.73-002149.6 &               & 7463 $\pm$  3 
&               & 8450 $\pm$ 20 & 9332 $\pm$ 31 \\ 
&                 &  68.0 $\pm$  1.7 &                 
&  13.9 $\pm$  2.2 &  28.6 $\pm$  6.1 \\ 
SDSSp J122600.68+005923.6 &               & 6210 $\pm$  0 
& 6846 $\pm$ 12 & 7369 $\pm$  8 & 8156 $\pm$  2 \\ 
&                 &  88.8 $\pm$  2.2 &   4.7 $\pm$  0.9 
 &   7.1 $\pm$  1.1 &  23.7 $\pm$  1.0 \\ 
SDSSp J123503.04-000331.8 & 5883 $\pm$  6 & 6927 $\pm$  1 
& 7434 $\pm$ 16 & 7960 $\pm$ 34 & 8807 $\pm$ 13 \\ 
&  31.3 $\pm$  6.1 &  85.8 $\pm$  8.1 &   6.7 $\pm$  2.2 
 &  10.6 $\pm$  5.1 &  29.4 $\pm$  5.0 \\  \\

SDSSp J131052.52-005533.4 &               & 6267 $\pm$  1 
&               &               & 7973 $\pm$  3 \\ 
&                 &  71.3 $\pm$  2.9 &                 
&                 &  28.5 $\pm$  1.2 \\ 
SDSSp J132110.82+003821.7 &               & 6967 $\pm$  2 
&               &               & 8830 $\pm$  5 \\ 
&                 &  42.8 $\pm$  5.6 &                 
&                 &  34.8 $\pm$  6.4 \\ 
SDSSp J141205.78-010152.6 &               & 5779 $\pm$  2 
& 6200 $\pm$ 12 & 6637 $\pm$ 13 & 7304 $\pm$ 14 \\ 
&                 &  79.7 $\pm$  4.3 &   2.0 $\pm$  0.9 
 &   8.1 $\pm$  1.3 &  13.6 $\pm$  2.5 \\ 
SDSSp J141332.35-004909.7 & 5315 $\pm$  2 &               
&               & 7197 $\pm$ 15 & 7971 $\pm$  2 \\ 
&  19.9 $\pm$  1.3 &                 &                 
&  11.7 $\pm$  1.9 &  28.3 $\pm$  1.4 \\ 
SDSSp J141315.36+000032.1 &               & 6183 $\pm$  1 
&               &               & 7875 $\pm$  4 \\ 
&                 &  69.8 $\pm$  4.1 &                 
&                 &  22.6 $\pm$  4.6 \\  \\

SDSSp J142329.98+004138.4 &               & 5843 $\pm$  4 
&               & 6679 $\pm$  9 & 7374 $\pm$  3 \\ 
&                 &  50.9 $\pm$  2.8 &                 
&  11.2 $\pm$  1.2 &  32.4 $\pm$  1.4 \\ 
SDSSp J142647.82+002740.4 & 4860 $\pm$  4 & 5727 $\pm$  1 
&               & 6388 $\pm$ 16 & 7265 $\pm$  4 \\ 
&  26.3 $\pm$  1.5 &  48.4 $\pm$  4.3 &                 
&  10.8 $\pm$  2.6 &  26.9 $\pm$  1.5 \\ 
SDSSp J144428.67-012344.1 &               & 6358 $\pm$  4 
&               & 7243 $\pm$ 26 & 7999 $\pm$ 12 \\ 
&                 &  76.3 $\pm$  4.9 &                 
&  13.3 $\pm$  5.1 &  37.2 $\pm$  2.9 \\ 
SDSSp J144713.08-012158.7 &               & 5897 $\pm$  3 
&               & 6736 $\pm$  7 & 7424 $\pm$  6 \\ 
&                 &  37.9 $\pm$  4.7 &                 
&  16.8 $\pm$  1.2 &  30.1 $\pm$  1.5 \\ 
SDSSp J151618.44-000544.3 &               & 5805 $\pm$ 12 
&               & 6576 $\pm$ 13 & 7299 $\pm$ 12 \\ 
&                 &  85.8 $\pm$ 16.2 &                 
&  34.3 $\pm$  3.3 &  42.3 $\pm$  3.4 \\  \\

SDSSp J152740.52-010602.6 &               & 6595 $\pm$  2 
&               & 7584 $\pm$  4 &              \\ 
&                 &  74.1 $\pm$  4.7 &                 
&   7.2 $\pm$  1.6 &                 \\
SDSSp J161926.87-011825.2 &               & 5891 $\pm$  3 
&               & 6792 $\pm$ 15 & 7491 $\pm$  5 \\ 
&                 &  69.0 $\pm$  6.2 &                 
&  15.7 $\pm$  1.9 &  36.2 $\pm$  1.8 \\ 
SDSSp J162116.91-004251.1 & 4853 $\pm$  6 & 5731 $\pm$  0 
& 6146 $\pm$  4 & 6589 $\pm$  6 & 7288 $\pm$  2 \\ 
&  19.4 $\pm$  2.0 &  82.3 $\pm$  1.6 &   6.2 $\pm$  0.6 
 &   7.2 $\pm$  0.7 &  43.8 $\pm$  1.0 \\ 
SDSSp J165527.61-000619.2 &               & 6061 $\pm$  1 
& 6512 $\pm$ 10 &               & 7740 $\pm$  4 \\ 
&                 & 172.0 $\pm$  6.0 &  14.6 $\pm$  3.0 
 &                 &  63.3 $\pm$  3.5 \\ 
\enddata
\tablenotetext{}{The two entries in each line are the central wavelength and rest frame equivalent width
from the Gaussian fit to the line profile, both measured in
\AA{}ngstroms.}
\tablenotetext{}{Emission line properties are not measured for
SDSSp J130348.94+002010.5, SDSSp J153259.96--003944.1  or SDSSp J160501.21--011220.0.}
\end{deluxetable}
\end{scriptsize}

\begin{deluxetable}{cccccccc}
\tablenum{4}
\tablecolumns{6}
\tablecaption{Continuum Properties of SDSS High-redshift Quasars}
\tablehead{quasar               & redshift   & E(B--V) & AB$_{1450}^a$ &$M_{B}$ &$
z_{LLS}$ & $z_{abs}^b$ }
 \startdata
SDSSp J$111401.48-005321.1$ & 4.58 $\pm$ 0.01 & 0.038 & 19.73 & --26.82  &4.50 &   \\ 
SDSSp J$112253.51+005329.8$ & 4.57 $\pm$ 0.02 & 0.036 & 19.20 & --27.35  &4.48 &   \\ 
SDSSp J$120441.73-002149.6$ & 5.03 $\pm$ 0.05 & 0.026 & 19.05 & --27.64  &     &   \\ 
SDSSp J$122600.68+005923.6$ & 4.25 $\pm$ 0.02 & 0.024 & 19.07 & --27.36  &4.10 &   \\ 
SDSSp J$123503.04-000331.8$ & 4.69 $\pm$ 0.01 & 0.022 & 20.30 & --26.29  &4.68 &   \\  \\

SDSSp J$130348.94+002010.4$ & 3.64 $\pm$ 0.03 & 0.020 & 18.88 & --27.31  &     &   \\
SDSSp J$131052.52-005533.4$ & 4.14 $\pm$ 0.01 & 0.025 & 18.93 & --27.46  &4.10 &   \\ 
SDSSp J$132110.82+003821.7$ & 4.70 $\pm$ 0.01 & 0.032 & 20.12 & --26.47  &4.59 &   \\ 
SDSSp J$141205.78-010152.6$ & 3.73 $\pm$ 0.02 & 0.057 & 19.32 & --26.91 &3.62 &3.26   \\ 
SDSSp J$141315.36+000032.1$ & 4.08 $\pm$ 0.02 & 0.048 & 20.07 & --26.30  &     &   \\  \\

SDSSp J$141332.35-004909.7$ & 4.14 $\pm$ 0.01 & 0.043 & 19.29 & --27.10  &     &   \\ 
SDSSp J$142329.98+004138.4$ & 3.76 $\pm$ 0.01& 0.027 & 19.75 & --26.49  &3.59 &   \\ 
SDSSp J$142647.82+002740.4$ & 3.69 $\pm$ 0.01& 0.031 & 19.53 & --26.69  &     &        \\ 
SDSSp J$144428.67-012344.1$ & 4.16 $\pm$ 0.01& 0.047 & 19.60 & --26.80  &4.06 &   \\ 
SDSSp J$144758.46-005055.4$ & 3.80 $\pm$ 0.01& 0.046 & 19.64 & --26.62  &3.53 &    \\ \\

SDSSp J$151618.44-000544.3$ & 3.70 $\pm$ 0.01& 0.057 & 20.10 & --26.11  &3.60 &3.03   \\ 
SDSSp J$152740.52-010602.6$ & 4.41 $\pm$ 0.02 & 0.156 & 19.62 & --26.87 &4.35 &   \\ 
SDSSp J$153259.96-003944.1$ & 4.62 $\pm$ 0.04 & 0.122 &  19.41 & --27.16  &4.62 & 4.58  \\ 
SDSSp J$160501.21-011220.0$ & 4.92 $\pm$ 0.05 & 0.199 & 19.43 & --27.23  &     &   \\ 
SDSSp J$161926.87-011825.2$ & 3.84 $\pm$ 0.01 & 0.119 & 19.73 & --26.54  &3.63 &   \\ \\

SDSSp J$162116.91-004251.1$ & 3.70 $\pm$ 0.01 & 0.098 & 17.41 & --28.81  &     &   \\ 
SDSSp J$165527.61-000619.2$ & 3.99 $\pm$ 0.01 & 0.307 & 20.01 & --26.33  &     &   \\ 
\enddata
\tablenotetext{}{Absolute magnitudes assume $H_0 = 50\rm\, km\, s^{-1}
\,Mpc^{-1}$, $q_0 = 0.5$ and power law index $\alpha=-0.5$.}
\tablenotetext{a}{The absolute flux calibration is accurate to $\sim 0.1$ mag.}
\tablenotetext{b}{These are redshifts of candidate damped Ly$\alpha$ lines seen in
the spectra.}
\end{deluxetable}

\newpage
\begin{figure}
\vspace{-3cm}

\epsfysize=400pt \epsfbox{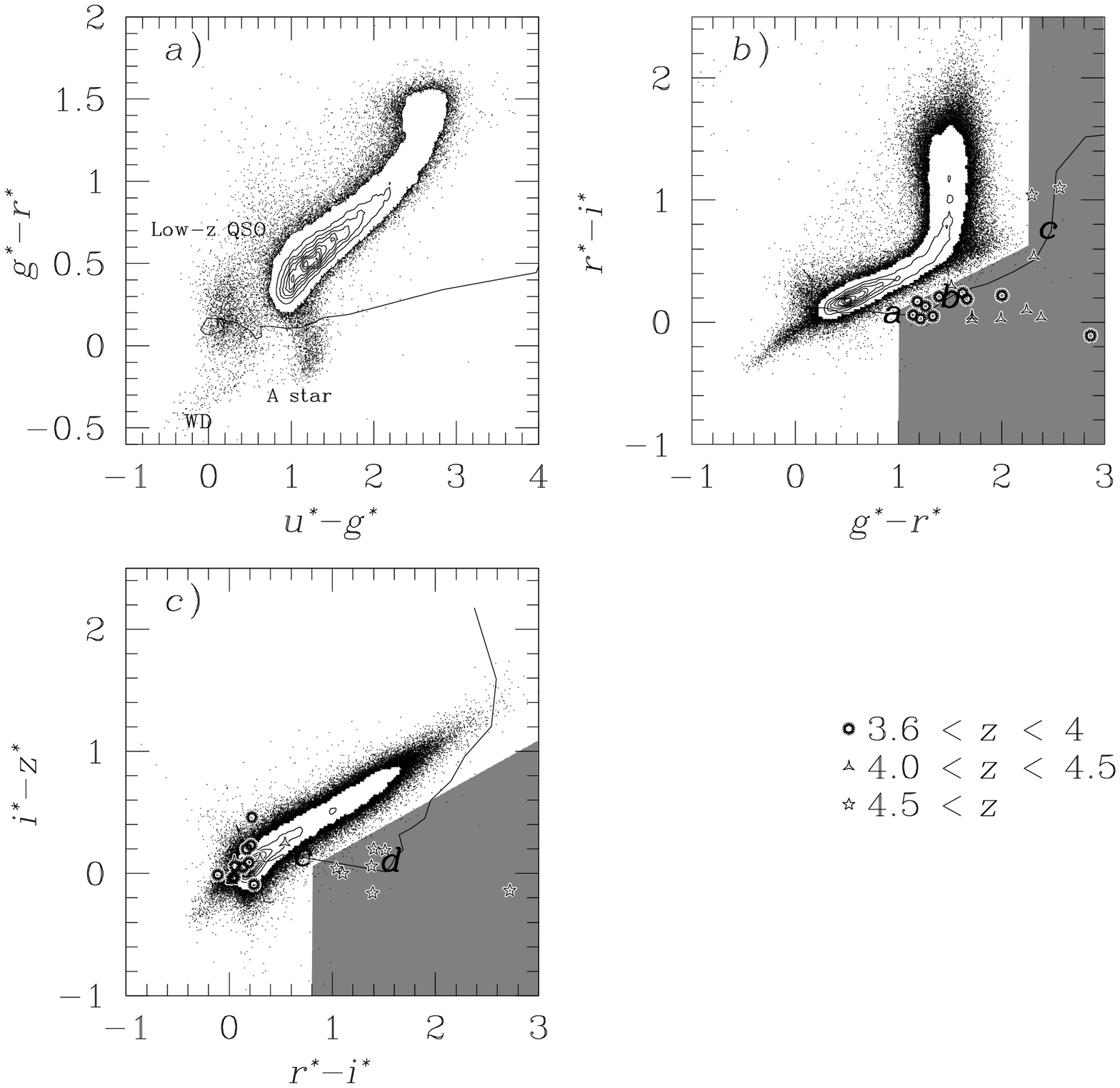}
\vspace{1cm}
Figure 1. Color-color diagrams for all stellar objects in 75 square
degrees of SDSS imaging data from Run 752, with $i^* < 20.2$.
The inner parts of the diagrams are shown as contours,
linearly spaced in the density of stars in color-color space.
The shaded areas on the $g^*- r^*$ vs. $r^* - i^*$ and
the $r^* - i^*$ vs. $i^* - z^*$ diagrams represent the
selection criteria used to select quasar candidates.
The solid line is the median track of simulated quasar colors
as a function of redshift (adapted from Fan 1999).  The letters {\it
a}, {\it b}, {\it c}, and {\it d} indicate the positions on the locus of median color quasars
at $z = 3.6, 4.1, 4.6$, and 5.0, respectively.
Colors of the 24 confirmed SDSS quasars at $z>3.6$ are also plotted
on the diagrams.
\end{figure}

\begin{figure}
\vspace{-0.5cm}

\epsfysize=500pt \epsfbox{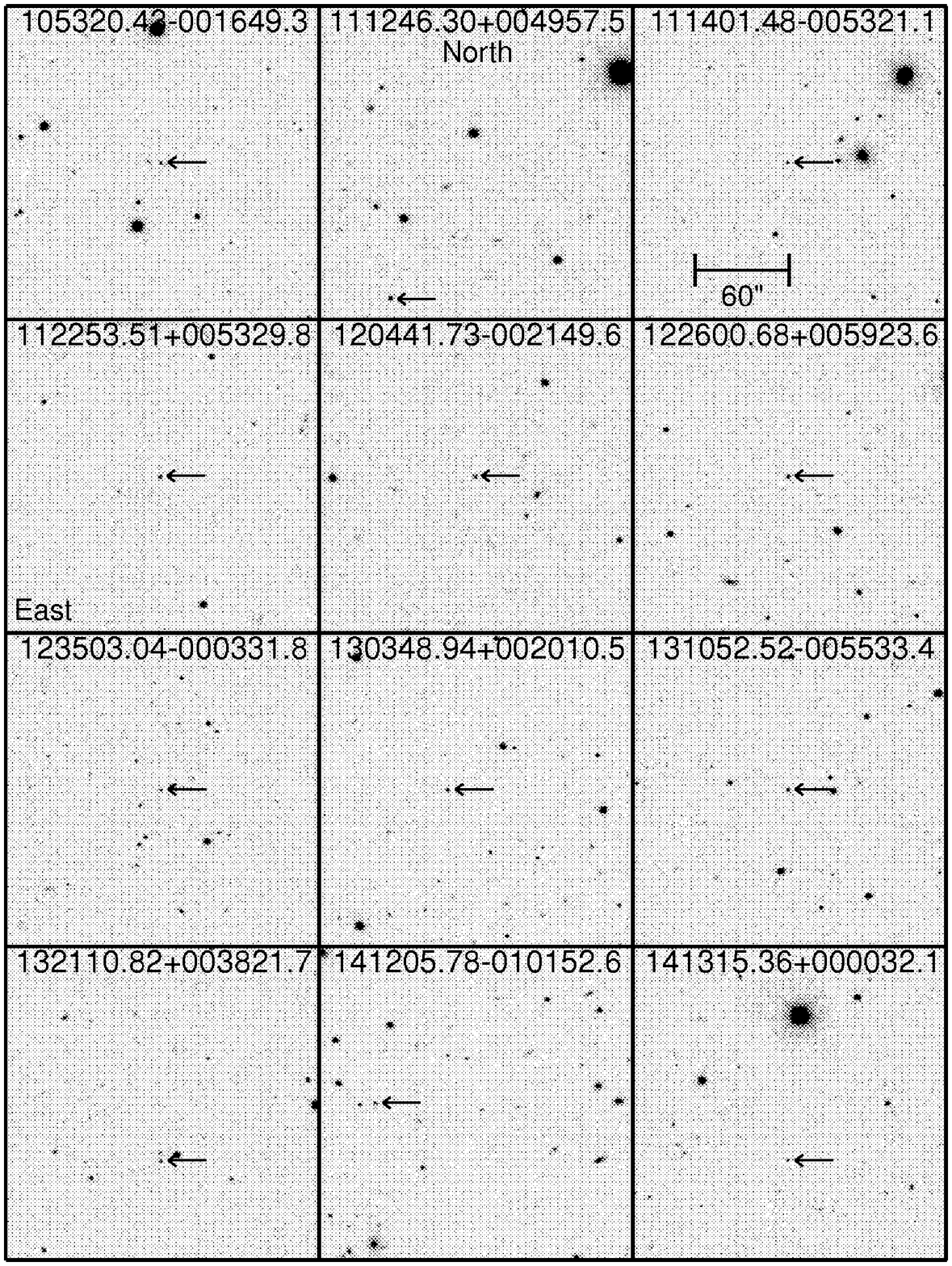}
\vspace{1cm}
Figure 2. Finding charts for the 24 SDSS quasars.
The data are $200'' \times 200''$ SDSS images in the $i'$ band
(54.1 sec exposure time). Most of them are re-constructed
from the atlas images and binned background from the SDSS database
(\cite{Lupton99b}). 
North is up; East is to the left.
\end{figure}

\begin{figure}
\vspace{-1cm}

\epsfysize=500pt \epsfbox{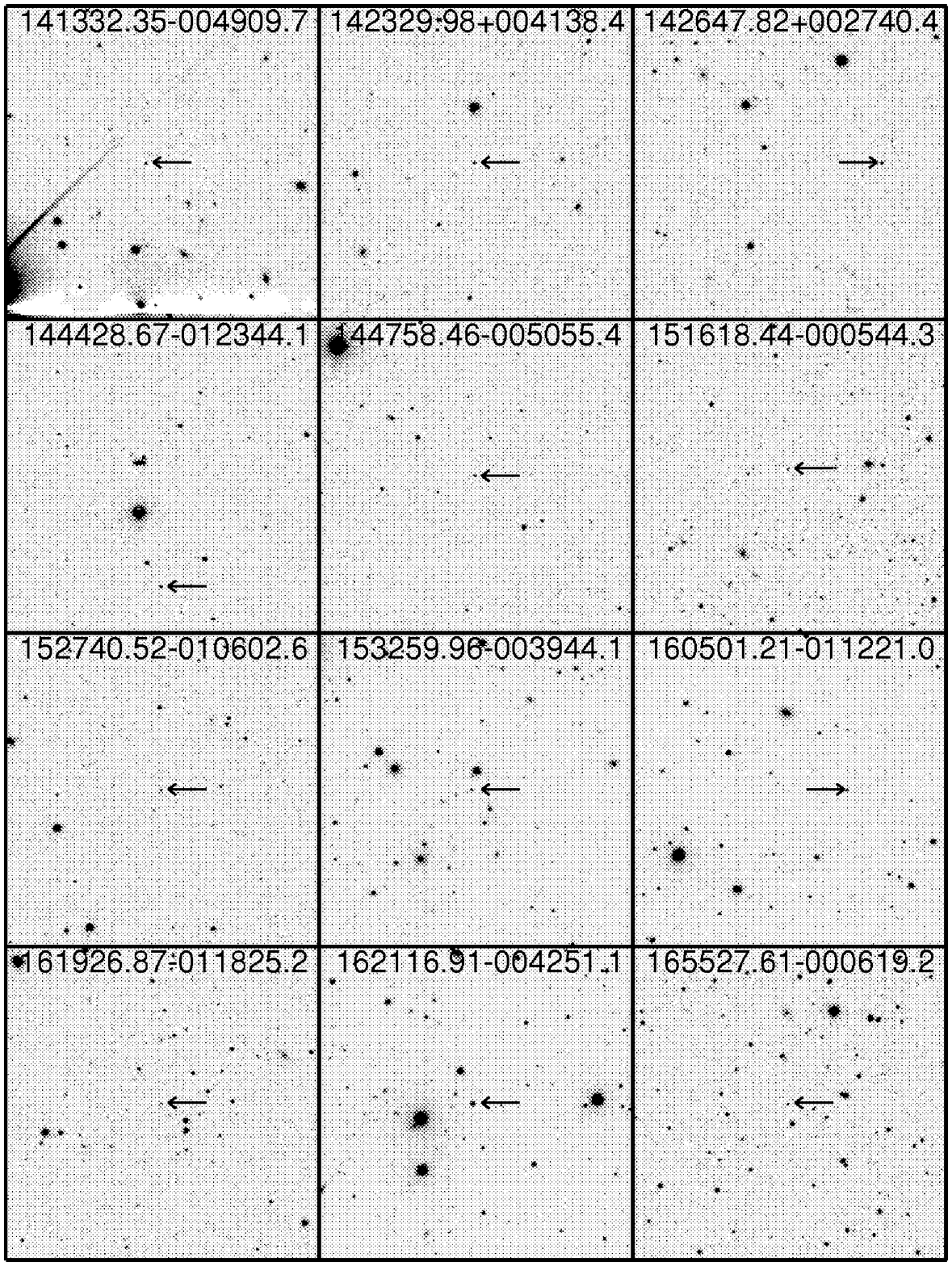}
\vspace{1cm}
Figure 2. Continued.
\end{figure}

\begin{figure}
\vspace{-3cm}

\epsfysize=600pt \epsfbox{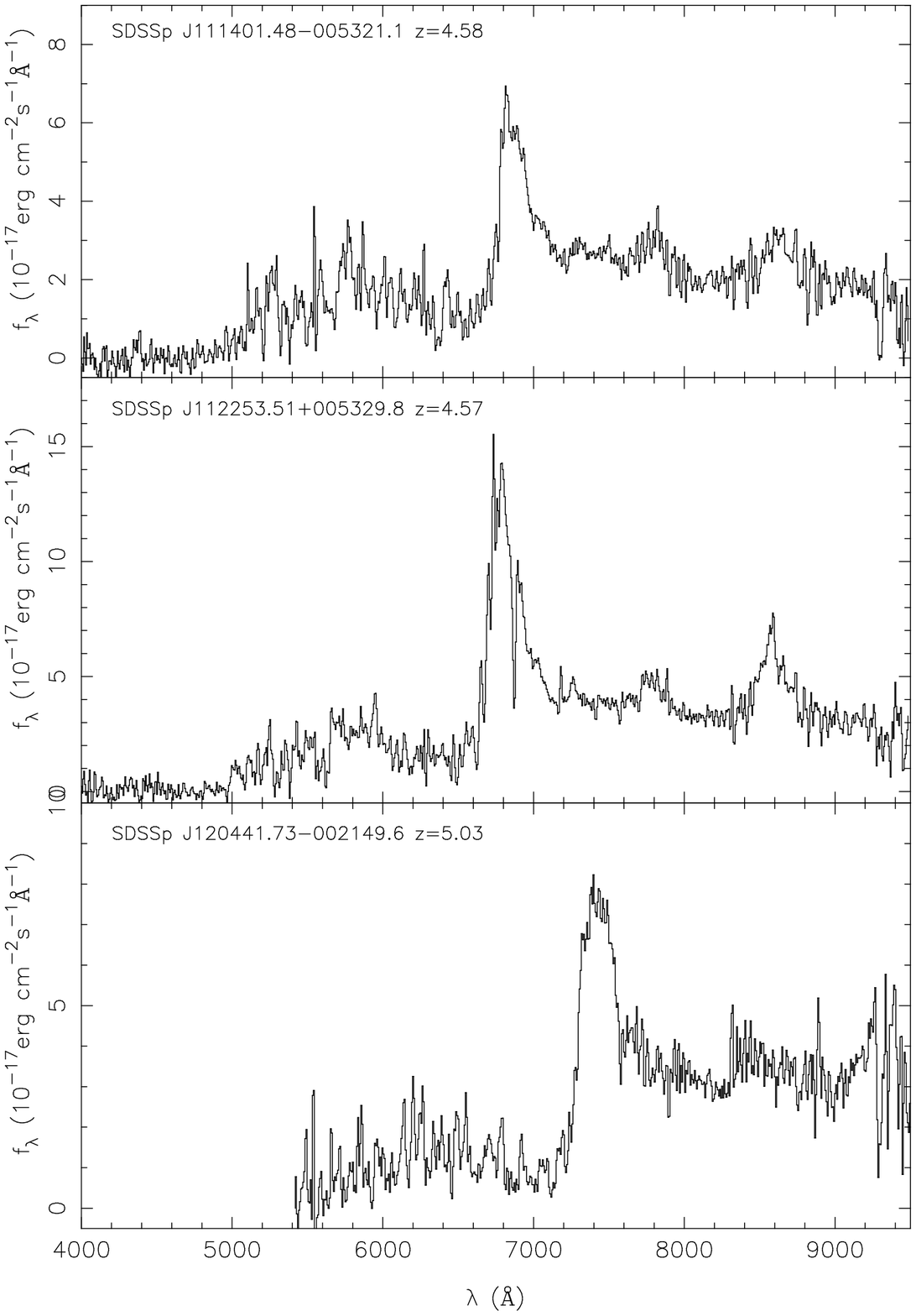}
\vspace{1cm}
Figure 3. ARC 3.5m/DIS spectra of 22 new SDSS quasars.
The spectral resolution is about 12 \AA\ in the blue and
14 \AA\ in the red.
Each pixel represents 6.2\AA.
Exposure times ranged from 600 sec to 5400 sec.
\end{figure}

\begin{figure}
\vspace{-3cm}

\epsfysize=600pt \epsfbox{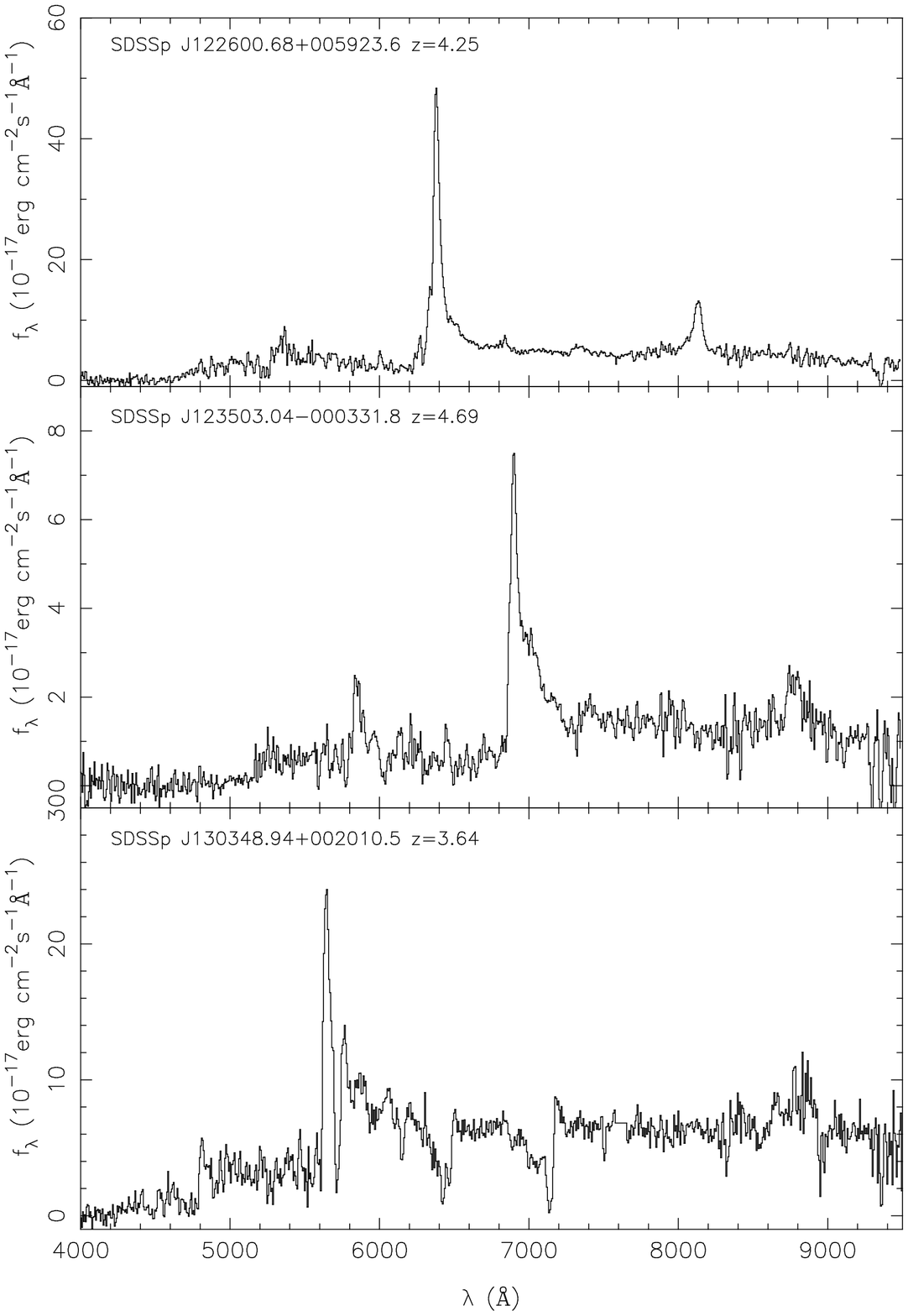}
\vspace{1cm}
Figure 3. Continued.
\end{figure}

\begin{figure}
\vspace{-3cm}

\epsfysize=600pt \epsfbox{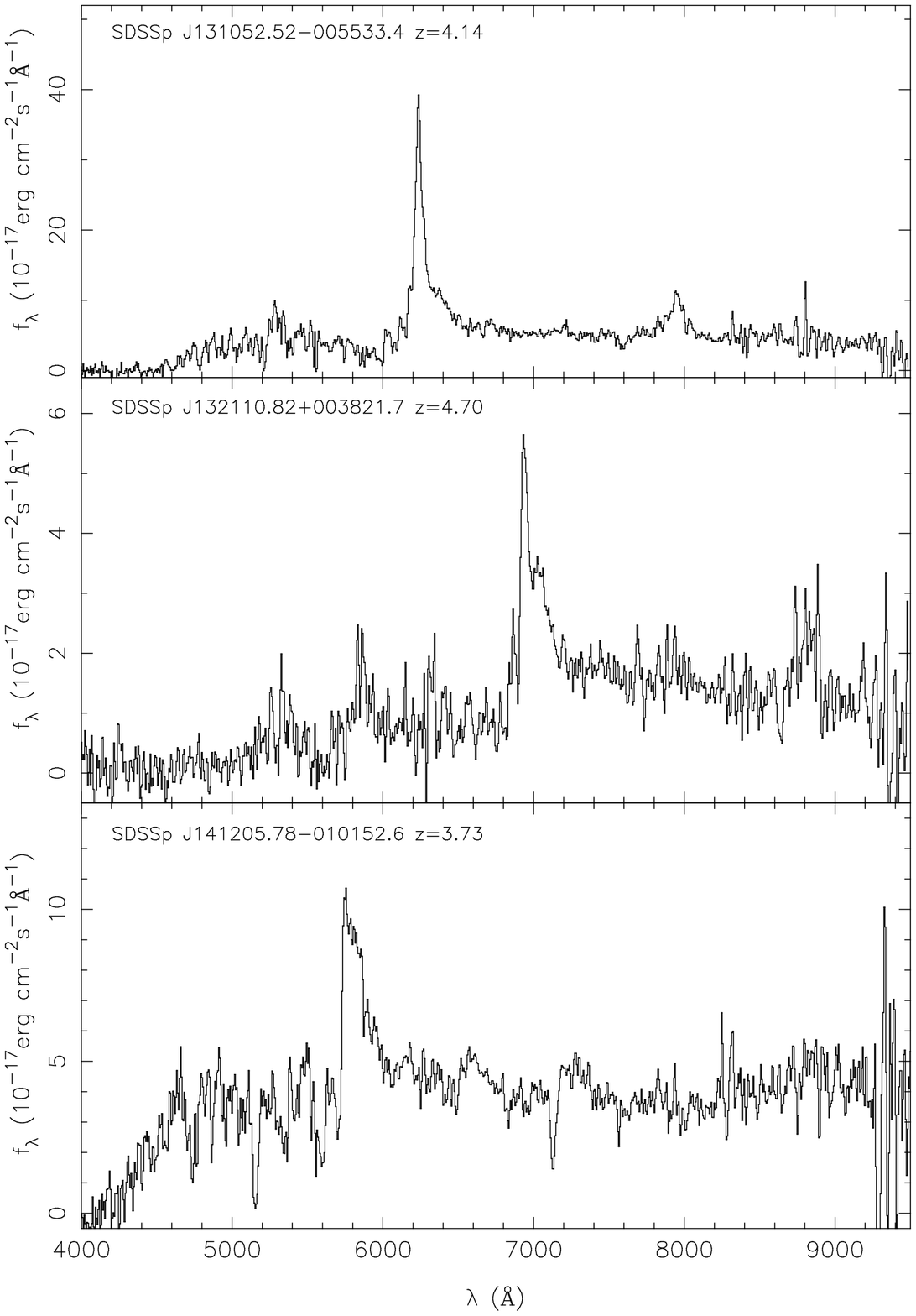}
\vspace{1cm}
Figure 3. Continued.
\end{figure}

\begin{figure}
\vspace{-3cm}

\epsfysize=600pt \epsfbox{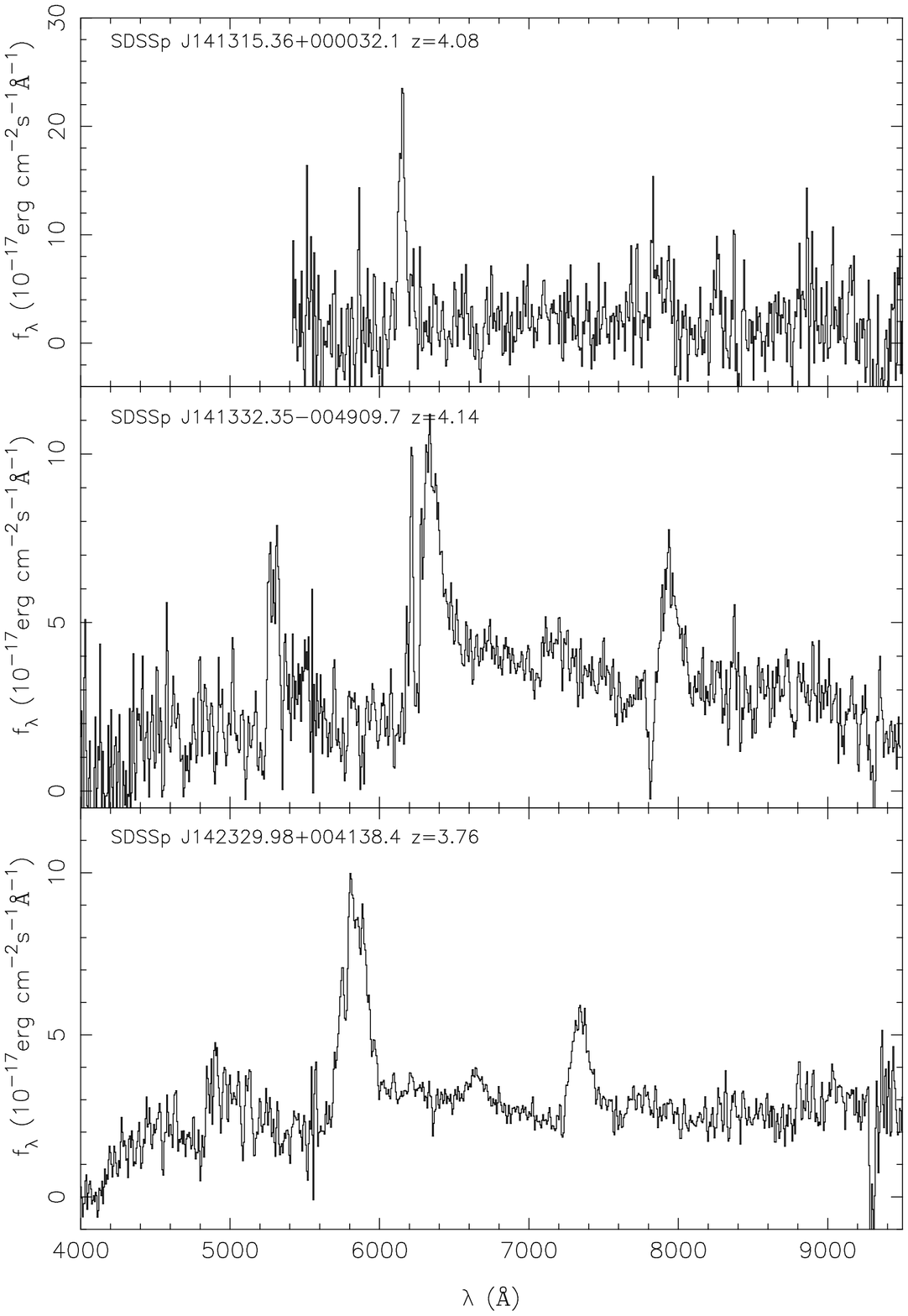}
\vspace{1cm}
Figure 3. Continued.
\end{figure}

\begin{figure}
\vspace{-3cm}

\epsfysize=600pt \epsfbox{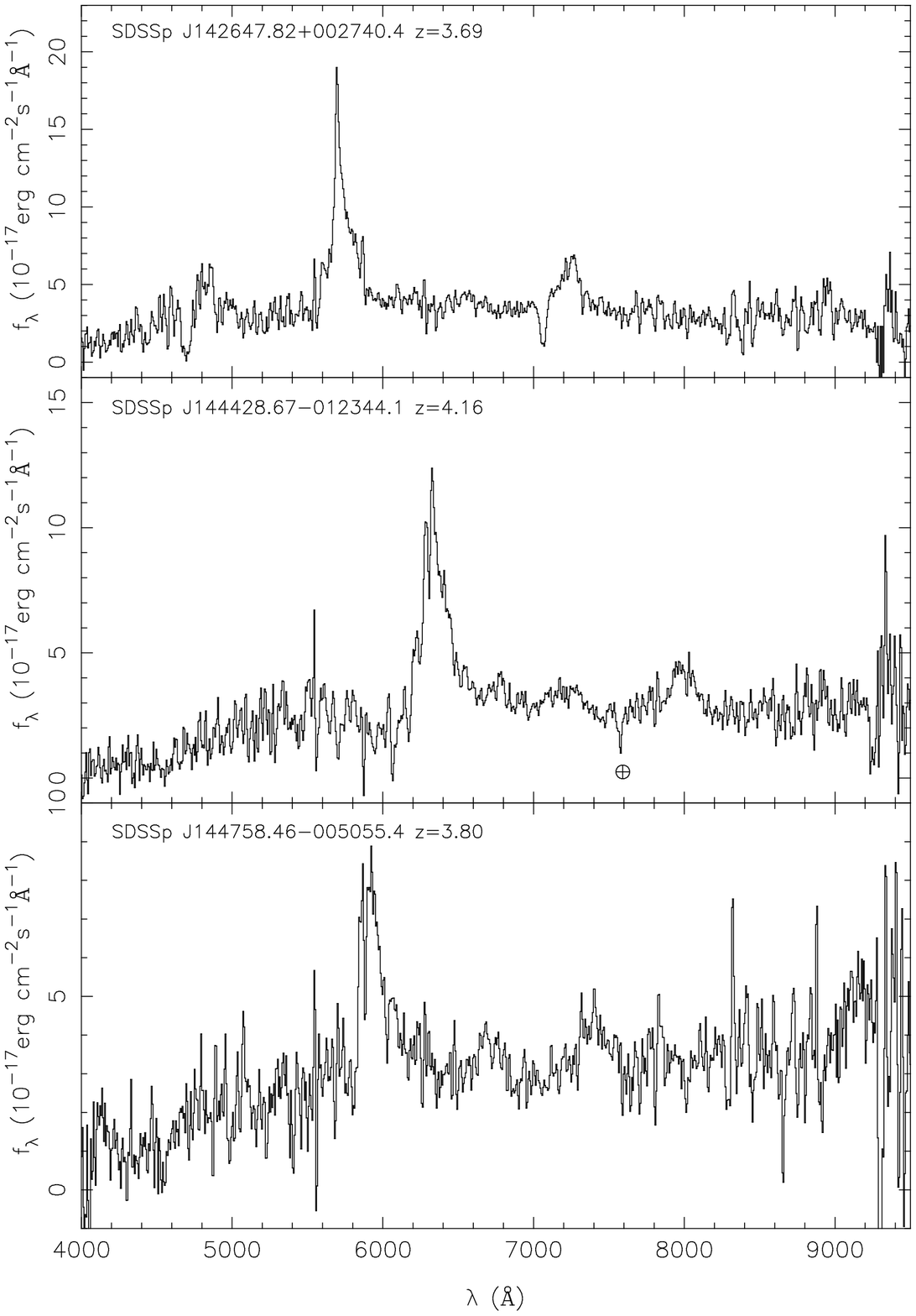}
\vspace{1cm}
Figure 3. Continued.
\end{figure}

\begin{figure}
\vspace{-3cm}

\epsfysize=600pt \epsfbox{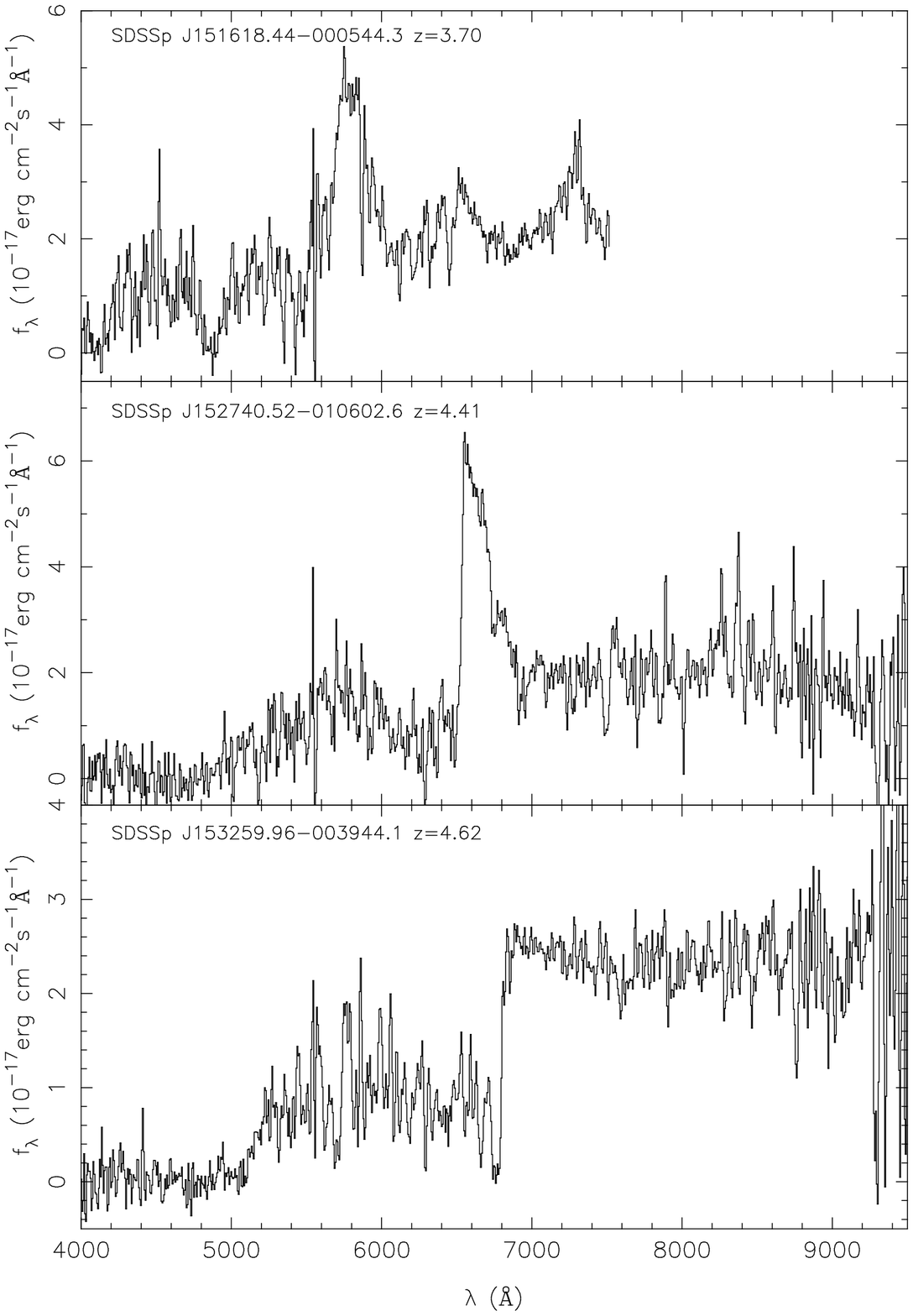}
\vspace{1cm}
Figure 3. Continued.
\end{figure}

\begin{figure}
\vspace{-3cm}

\epsfysize=600pt \epsfbox{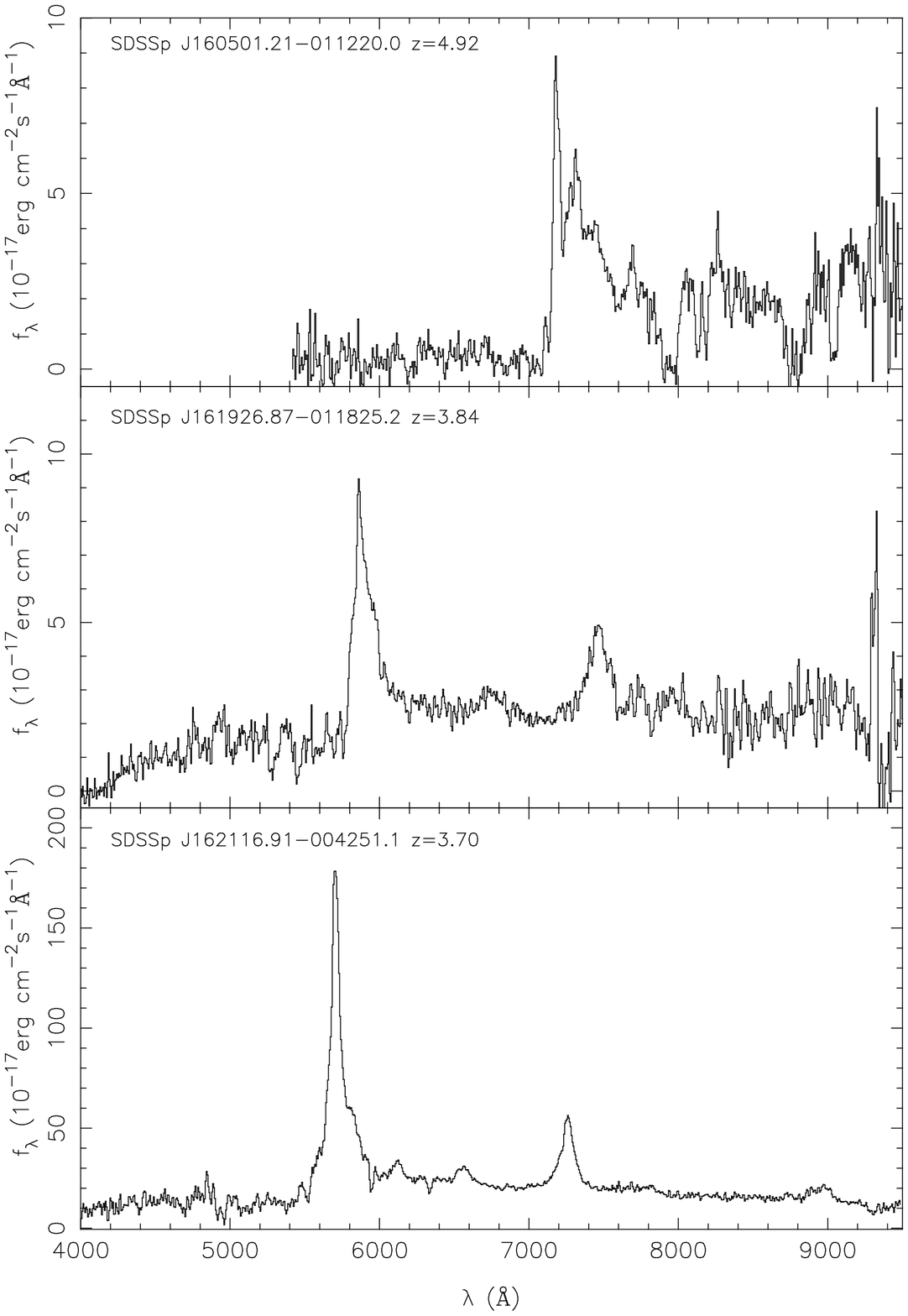}
\vspace{1cm}
Figure 3. Continued.
\end{figure}

\begin{figure}
\vspace{-3cm}

\epsfysize=600pt \epsfbox{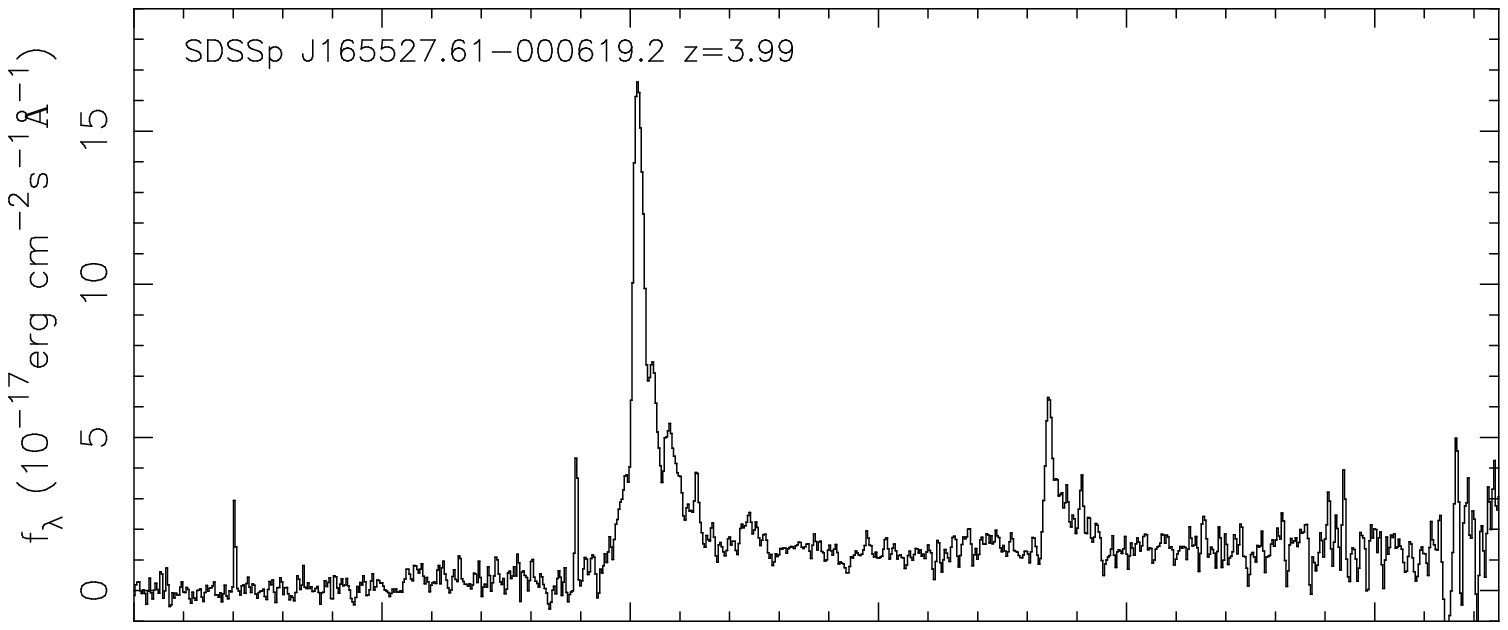}
\vspace{1cm}
Figure 3. Continued.
\end{figure}

\begin{figure}
\vspace{-2cm}

\epsfysize=600pt \epsfbox{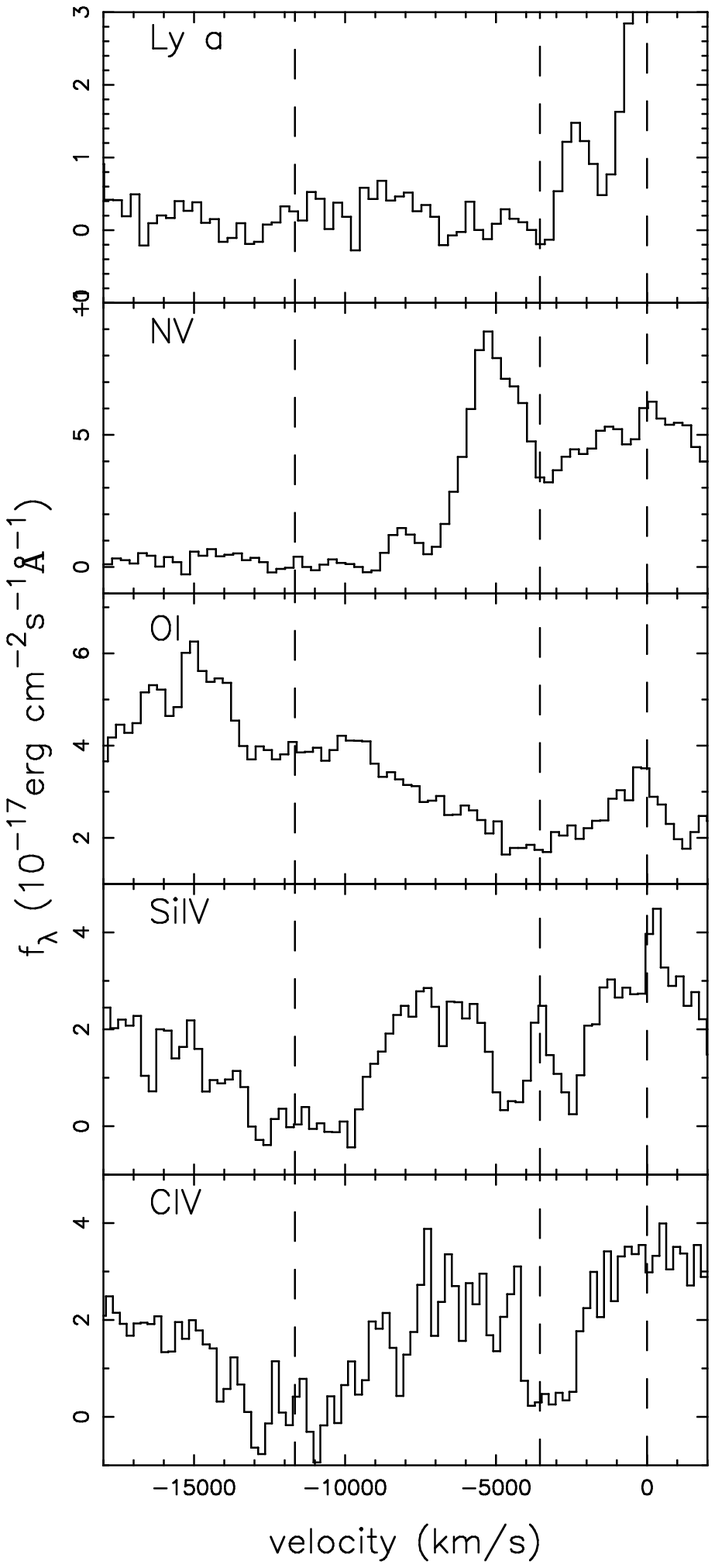}
\vspace{1cm}
Figure 4. The Broad Absorption Lines of SDSSp J160501.21--011220.0.
Absorption lines of Ly$\alpha$, N$\,$V, O$\,$I, Si$\,$IV and C$\,$IV
are shown, although both systems are not seen in every line.
The spectra are aligned in velocity space relative to the
emission line redshift of the quasar ($z=4.92$).
The velocity resolution is about 500 $\rm \ km\ s^{-1}$.
Two BALs are seen, one at $z=4.86$ ($v=-3000\rm \ km\ s^{-1}$), 
and the other at $z=4.69$ ($v=-11700\rm \ km\ s^{-1}$).
\end{figure}
\end{document}